\documentclass[journal=jctcce,manuscript=article]{achemso}

\usepackage{natbib}

\usepackage{amssymb}
\usepackage{amsmath}
\usepackage{amsfonts}
\usepackage{graphicx}

\newcommand{\rcite}[1]{Ref.~\citenum{#1}}
\newcommand{\rcites}[1]{Refs.~\citenum{#1}}

\renewcommand{\vec}[1]{\boldsymbol{#1}}

\newcommand{\beq}{\begin{eqnarray}}
\newcommand{\eeq}{\end{eqnarray}}

\newcommand{\half}{\frac{1}{2}}

\newcommand{\Hxc}{\text{Hxc}}
\newcommand{\xrm}{\text{x}}

\newcommand{\Hrm}{\text{H}}

\newcommand{\Ext}{{\text{ext}}}

\newcommand{\pr}{^{\prime}}

\newcommand{\vr}{\vec{r}}

\newcommand{\vrp}{\vec{r}\pr}

\newcommand{\ibraketop}[3]{\langle#1|#2|#3\rangle}

\newcommand{\up}{\uparrow}
\newcommand{\down}{\downarrow}

\title{$C_6$ coefficients and dipole polarizabilities for all
atoms and many ions in rows 1-6 of the periodic table}
\author{Tim Gould}
\email{t.gould@griffith.edu.au}
\affiliation{Qld Micro- and Nanotechnology Centre, %
Griffith University, Nathan, Qld 4111, Australia}
\author{{Tom\'a\v{s} Bu\v{c}ko}}
\email{bucko@fns.uniba.sk}
\affiliation{Department of Physical and Theoretical Chemistry, Faculty of Natural Sciences, Comenius University in Bratislava, Mlynsk\'{a} Dolina, Ilkovi\v{c}ova 6, SK-84215 Bratislava, SLOVAKIA}
\affiliation{Institute of Inorganic Chemistry, Slovak Academy of Sciences, D\'ubravsk\'a cesta 9, SK-84236 Bratislava, SLOVAKIA} 
\begin{document}

%\begin{tocentry}
%  \includegraphics[width=\linewidth]{Figs/TOC_Image}
%\end{tocentry}

\begin{abstract}
  Using time-dependent density functional theory (TDDFT) with
  exchange kernels we calculate and test imaginary frequency-dependent
  dipole polarizabilities for all atoms and many ions in rows 1-6
  of the periodic table. These are then integrated over
  frequency to produce $C_6$ coefficients.
  Results are presented under different models:
  straight TDDFT calculations using two different kernels;
  ``benchmark'' TDDFT calculations corrected by more accurate
  quantum chemical and experimental data;
  and ``benchmark'' TDDFT with frozen orbital anions.
  Parametrisations are presented for 411+ atoms and ions, allowing
  results to be easily used by other researchers. 
  A curious relationship, $C_{6,XY}\propto [\alpha_X(0)\alpha_Y(0)]^{0.73}$
  is found between $C_6$ coefficients and static
  polarizabilities $\alpha(0)$. The relationship
  $C_{6,XY}=2C_{6,X}C_{6,Y}/[\alpha_X/\alpha_YC_{6,Y}+\alpha_Y/\alpha_XC_{6,X}]$
  is tested and found to work well ($<5$\% errors) in about
  80\% of cases, but can break down badly ($>30$\% errors) in
  a small fraction of cases.
\end{abstract}

The importance of van der Waals (vdW) forces in physical systems,
especially at the nanoscale, is increasingly being recognised
(see e.g. \rcites{Dobson2012-JPCM,Eshuis2012,Ambrosetti2016}
and references therein). This renewed interest has come about
in part because vdW forces are so vital to binding in layered
materials such as graphene\cite{Geim2013,Lebegue2010,%
Bjorkman2012,Bjorkman2012-Review,Bjorkman2014,Gobre2013}.
Unfortunately, conventional electronic structure techniques like
density functional theory in its common approximations
(e.g. GGAs\cite{GGA})
do not reproduce van der Waals forces\cite{Dobson2001}
which can lead to poor predictions for important
systems\cite{Bjorkman2012}.
To remedy this lack, a cornucopia of new approaches has
been developed over the past twenty years\cite{DobsonDinte,
Lundqvist1995,ALL,Dion2004,Langreth2005,
Grimme2004,Grimme2006,Grimme2010,
Tkatchenko2009,Tkatchenko2012,DiStasio2012,Dobson2012-JPCM,Eshuis2012}
that allow van der Waals forces to be included alongside
more conventional density functional
approximations\cite{Becke1988,GGA,LYP}.

A number of these approaches - most notably those based around
the work of Grimme\cite{Grimme2004,Grimme2006,Grimme2010}
and Tkatchenko and Scheffler\cite{Tkatchenko2009,Tkatchenko2012,%
DiStasio2012,Bucko2014-Iterative,Bucko2016-MBD,Ambrosetti2016}
- are based on atom-in-molecule (AIM) approximations. These
(arguably) semi-empirical approximations involve taking free atomic
$C_6$ coefficients and using them
to determine binding in more complex bulk and molecular systems.
These methods have achieved a great many successes
(including in some complex systems\cite{Gobre2013}).
Although it should be noted that they also have
flaws\cite{Bjorkman2012-Review,Dobson2014-IJQC,Bucko2014-Iterative}.

A vital componment of these AIM approaches is the data set of
$C_6$ coefficients. These must be pre-calculated elsewhere and
tabulated for input into the AIM scheme. However, determining
accurate $C_6$ coefficients is difficult. It requires as input
the dipole polarizability of a system across a range of
(imaginary) frequencies which can then be integrated
[see Eq.~\eqref{eqn:CP}]
to determine the dipole-dipole $C_6$ coefficient.
While static polarizabilities of closed- and some open-shell
atoms and small molecules can be calculated very accurately
(see e.g. \rcites{Mitroy2010,Schwerdtfeger2006} and references therein)
using sophisticated quantum chemical techniques, the dynamic
polarizabilities are more difficult to evaluate.
Reactive open-shell atoms and ions, which dominate chemistry
and materials science, represent a particular challenge for such
methods. Thus AIM inputs are often least reliable for the
elements that play the most interesting role in materials.

One route around the limitations of quantum chemcial
approaches is to employ time-dependent density functional
theory (TDDFT).
In linear response TDDFT the dipole polarizability
is obtained directly from the density-density response of an
atomic or ionic system, making its evaluation relatively
straightforward. Although they are not as efficient as regular DFT
calculations, TDDFT calculations are generally more accurate than
DFT and more efficient than
accurate quantum chemical calculations. They thus provide a middle
ground between speed and accuracy, allowing fairly accurate
calculations to be carried our in reasonable time even for systems
that are essentially intractable for more accurate quantum
chemical approaches.

Chu and Dalgarno, in their 2004 paper\cite{Chu2004},
took advantage of the ``balanced'' nature
(in terms of numerical cost and accuracy)
of TDDFT to determine accurate $C_6$ coefficients for open-shell
atoms, and were able to provide a complete set of coefficients
for elements in rows 1-3 of the periodic table.
Grimme \latin{et al} used a similar approach\cite{Grimme2010}
to determine atomic coefficients for all atoms in Rows 1-7.
But ``benchmark'' quality results for open shell elements
in rows~4-7 remain elusive. Open-shell ionic coefficients
are even less well studied, even in the upper rows. 

This manuscript thus seeks to use TDDFT calculations, loosely
based on the basic approach of Chu and Dalgarno, to provide
frequency-dependent dipole polarizabilities for all
atoms and interesting ions in rows~1-6 of the periodic table.
It will further present the results in an easy to use
parametrised model, making results readily available for use
in AIM calculations.
\footnote{These techniques could, in principle, be employed for row~7
too. But numerical issues and strong relativistic contributions
make results for these elements likely to be inaccurate. We thus
focus on reliable results for the more common, higher row elements.}

\section{Theory}

The well-known van der Waals (or dispersion) $C_6$ coefficient
usually appears in the formula $U_{\text{vdW}}=-C_{6,XY}/D_{XY}^6$ governing
a long-range attractive potential between well-separated,
localised systems $X$ and $Y$. For spherically symmetric systems,
in which the dipole polarizability tensor is a scalar times
the identity tensor,
$C_6$ can be found through the simplified Casimir-Polder formula
\begin{align}
  C_{6,XY}=&\frac{3}{\pi}\int d\omega \alpha_X(i\omega)\alpha_Y(i\omega),
  \label{eqn:CP}
\end{align}
which can be derived from many-body perturbation theory on the
Coulomb interaction between the two systems. Here
the $C_6$ coefficient depends on the frequency dependent
dipole polarizability $\alpha_{X/Y}(i\omega)$, evaluated at
imaginary frequency $i\omega$. Thus, knowing
$\alpha_X(i\omega)$ for a number of atoms and ions $X$
is sufficient to calculate the $C_6$ interaction between all
possible pairs. Furthermore, $\alpha(i\omega)$ can be used to calculate
higher order ``non-additive'' interactions\cite{Dobson2014-IJQC},
such as the Axilrod-Teller interaction\cite{AxilrodTeller}
or higher order dipolar contributions\cite{DiStasio2012}.
Note we use atomic units $\hbar=e^2/(4\epsilon_0)=m_e=1$
(giving energies in Ha and lengths in
Bohr radii $a_0$) in equation \eqref{eqn:CP} and
in all subsequent equations.

Clearly these dipolar dispersion forces are not the only
contribution to the long-range force between such systems.
Interactions between free ions are dominated by
forces from the Coulomb potential $U_C=q_Xq_Y/D_{XY}$.
Additionally, higher order terms, such as $C_8$ coefficients
in $U_8=-C_{8,XY}/D_{XY}^8$ also contribute, and these
must be obtained from quadropolar and higher interactions.
But the dipolar force plays an important role in all systems,
especially in embedded ions where Coulomb forces cancel
out over a molecule, leaving induction (which depends on $\alpha(0)$)
and dispersion forces as the leading long-range force terms.

The relationship \eqref{eqn:CP} between polarizabilities and
dispersion coefficients has been exploited
indirectly in various methods%
\cite{Grimme2004,Grimme2006,Grimme2010,Tkatchenko2009,Tkatchenko2012,%
DiStasio2012,Bucko2014-Iterative,Bucko2016-MBD,Ambrosetti2016}.
But it has been more rarely (if ever) exploited directly in
atom-in-molecule approaches. In part this is likely due to a
lack of viable frequency-dependent data for atoms and ions.
We will thus outline, in the remainder of this theory section,
how to calculate polarizabilities that can be used directly
in van der Waals calculations
using \eqref{eqn:CP} or related formulae.

\subsection{TDDFT Methodology}
\label{sec:Method}

At the heart of our results are linear response TDDFT
calculations using the \emph{all electron} numerical method
described in \rcites{Gould2012-RXH,Gould2012-2,%
Gould2013-LEXX,Gould2013-Aff}.
In summary, we calculate linear response functions using
time-dependent density functional theory in an ensemble averaged
(where appropriate) exact exchange groundstate\cite{Gould2013-LEXX}
and using an equivalent ensemble averaged\cite{Gould2012-RXH}
version of the Petersilka, Gossman and Gross\cite{PGG}
kernel. The radial exchange hole kernel\cite{Gould2012-RXH}
is also employed for a further test of accuracy.
We do not explictly include any relativistic calculations.
However, for our final benchmarks we do include them implicitly
via corrections based on accurate reference static
dipole polarizabilities (see Sec.~\ref{sec:Correction}).

To calculate polarizabilities we employ a post-DFT linear
response formalism. In this approach, we first calculate
the groundstate of the system using a DFT approximation.
Once the groundstate properties are calculated, we use them
to determine the response function $\chi$ governing the small
change in densities to a small change to the potential
of form $\Delta v(\vr;t)=\Delta v(\vr) e^{\omega t}$.%
\footnote{Conventionally $\chi$ is defined as the
response to periodic $e^{-i\omega t}$. Thus we refer
to the response to imaginary frequency $i\omega$ when
considering the real exponential $e^{\omega t}$.}
From $\chi$ we can calculate the spherically
averaged dipole polarizability using
\begin{align}
  \alpha(i\omega)=&\int d\vr d\vrp zz'
  \chi(\vr,\vrp;i\omega)
  \label{eqn:alpha}
\end{align}
and, if desired, $C_6$ coefficients can be evaluated using the
Casimir-Polder formula \eqref{eqn:CP}.
We note that $\alpha$ depends on electron number $N$ and nuclear
charge $Z$. In our calculations, the dependence on $N$
is via the occupation factors, while the dependence on $Z$
is via the external potential $v_{\Ext}(r)=-Z/r$.

In our calculation we start from groundstate Kohn-Sham properties
[notably potentials $v_s(\vr)$, occupation factors $f_i$,
orbitals $\phi_i(\vr)$ and densities $n(\vr)$] calculated
in ensemble DFT. The use of ensembles allows us to properly
account for atomic symmetries despite working from
spherical and spin-unresolved groundstates in which
$n(\vr)=n(r)$ and $n_{\up}(r)=n_{\down}(r)=n(r)/2$.
We employ the LEXX approximation\cite{Gould2013-LEXX}
adapted to general open shell systems with $d$ and $f$ orbitals.
For atoms and ions in Rows 4-6 we make a further simplifying
approximation: that the shells fill ``trivially'' according
to Hund's rules. This can lead to energies that are higher
than other fillings, but for reasons discussed in Appendix~\ref{app:TM}
we feel they are a more appropriate starting point.

In LEXX theory, the groundstate energy is approximated by
an orbital energy functional
\begin{align}
  E_0[n]=&\sum_i f_it_{s,i} + \int v_{\Ext}(r)n(r) d\vr
  \nonumber\\&
  +\sum_{ij}(F_{ij}^{\Hrm}P_{ij} - \half F_{ij}^{\xrm}Q_{ij})
  \label{eqn:ELEXX}
\end{align}
where the pair occupation factors $F_{ij}$ are determined by the
orbital occupation factors $f_i$. Here
$\sum_i f_i=N$ is the total number of electrons.
The energy terms are
\begin{align}
  t_{s,i}=&-\half\int d\vr \nabla^2\rho_i(\vr,\vrp)|_{\vr=\vrp},
  \\
  P_{ij}=&\int \frac{d\vr d\vrp}{2|\vr-\vrp|}n_i(\vr)n_j(\vrp),
  \\
  Q_{ij}=&\int \frac{d\vr d\vrp}{2|\vr-\vrp|}
  \rho_i(\vr,\vrp)\rho_j(\vrp,\vr),
\end{align}
where $\rho_i(\vr,\vrp)=\phi_i^*(\vr)\phi_i(\vrp)$ and
$n_i(\vr)=\rho_i(\vr,\vr)$. In cases where all orbitals are equally
occupied (i.e. where $f_i=2$ or $0$ for all orbitals)
$F_{ij}=f_if_j$ and \eqref{eqn:ELEXX}
is identical to conventional EXX theory.

The Kohn-Sham potential $v_s$ is found using the
Krieger, Li and Iafrate\cite{KLI1992}
(KLI) approximation to the optimized effective potential.
Here we write
\begin{align}
  v_s(\vr)=&v_{\Ext}(\vr)
  + v_{\Hxc}[\{\phi_i\},\{f_i,F_{ij}^{\Hrm},F_{ij}^{\xrm}\}](\vr)
\end{align}
and determine $v_{\Hxc}$ using the orbitals and occupation factors.
The orbitals themselves obey
\begin{align}
  [-\half\nabla^2 + v_s(\vr) - \epsilon_i]\phi_i(\vr)=&0
  \label{eqn:Ham}
\end{align}
so that we need to iterate to self-consistency.
LEXX+KLI should (and we have found no evidence to the
contrary) yield good approximations for the potential, orbital
and density of atoms, at least up to relativistic effects.

To calculate the polarizability we need the density response
function $\chi$ from linear-response\cite{Zaremba1976}
time dependent DFT (TDDFT). To find $\chi$ we solve
\begin{align}
  \chi(\vr,\vrp;i\omega)=&\chi_0 + \chi_0\star f_{\Hxc} \star \chi,
  \label{eqn:chi}
\end{align}
where the spin-symmetry allows us to ignore spin.
Here the convolution $\star$ indicates an integral
over the interior space variable
such that $[f\star g](\vr,\vrp)=\int d\vr_2 f(\vr,\vr_2)g(\vr_2,\vrp)$.

Equation \eqref{eqn:chi} requires two inputs.
Firstly, it needs the non-interacting response function
$\chi_0(\vr,\vrp;i\omega)$, which governs the change in
density at $\vr$
in response to changes in the Kohn-Sham potential $v_s$
at $\vrp$.
For our calculations, we determine $\chi_0$ using
\begin{align}
  \chi_0(\vr,\vrp;i\omega)=&2\Re\sum_i f_i\rho_i(\vr,\vrp)
  G(\vr,\vrp;\epsilon_i-i\omega).
  \label{eqn:chi0}
\end{align}
Here the KS Greens function
$G(\vr,\vrp;\epsilon_i-i\omega)
=-\sum_j \frac{\rho_i(\vrp,\vr)}{\epsilon_j-\epsilon_i+i\omega}$
obeys
\begin{align}
  [-\half\nabla^2 + v_s - \Omega]G(\vr,\vrp;\Omega)=&-\delta(\vr-\vrp).
  \label{eqn:HamG}
\end{align}
In our implementation we use the effective one dimensionality
of the Hamiltonian to solve $G$ directly using \eqref{eqn:HamG}
rather than using the sum form.
This allows us to calculate very accurate Greens functions using
a shooting method and the cusp condition.

Secondly, we need to evaluate the Hartree, exchange and
correlation kernel
\begin{align}
  f_{\Hxc}(\vr,\vrp;i\omega)
  =&\frac{\delta v_{\Hxc}(\vr)}{\delta n(\vrp)}|_{i\omega}
  =\frac{\delta v_{\Hxc}(\vrp)}{\delta n(\vr)}|_{i\omega},
\end{align}
which is not known exactly and must be approximated.
In this work we employ two different approximations
for the kernel $f_{\Hxc}$. For most calculations we use the
Petersilka, Gossman, Gross\cite{PGG} approximation to the
kernel adapted for LEXX. The resulting kernel takes the form
\begin{align}
  f_{\Hxc}^{\text{PGG}}=&
  \frac{\sum_{ij}[F_{ij}^{\Hrm}n_i(\vr)n_j(\vrp)
      -\half F_{ij}^{\xrm}\rho_i(\vr,\vrp)\rho_j(\vrp,\vr)]}{n(\vr)n(\vrp)}.
\end{align}
For additional tests
we also employ the radial exchange hole (RXH)
kernel\cite{Gould2012-RXH} similarly adapted to open-shell atoms.
Both kernels are more accurate than the popular
random phase approximation (RPA) due to their inclusion of
dynamic screening effects and consequent reduction in
self interaction errors
(see e.g. \rcites{Gould2012-RXH,Gould2013-Aff}).

Finally, we can calculate $\chi(\vr,\vr';i\omega)$ for a given
$N$ and $Z$ using \eqref{eqn:chi} and its inputs.
$\alpha(i\omega)$ follows from \eqref{eqn:alpha}.
More formally, we calculate the response
$\chi[v_s,\{f_i\},f_{\Hxc}](\vr,\vr';i\omega)$,
as a functional of any spherically symmetric effective potential
$v_s$, any set of occupation factors $\{f_i\}$ obeying $\sum_if_i=N$,
and any kernel $f_{\Hxc}$.
In the two approximations presented here,
$f_{\Hxc}$ is itself entirely determined by $v_s$ and $\{f_i\}$,
allowing us to drop the dependence on the kernel.

\subsubsection{Numerical convergence}
\label{sec:converge}

All calculations are carried out using a bespoke
radial code which, together with the LEXX\cite{Gould2013-LEXX}
approximation, is designed to perform highly accurate calculations
without problems with basis set convergence. The code makes use of
the spherical symmetry of the Kohn-Sham potential
and employs a radial grid to calculate orbitals and Greens
functions using a shooting method. Details of the method are
described in
\rcites{Gould2012-RXH,Gould2012-2,Gould2013-LEXX,Gould2013-Aff},
together with some details on convergence and accuracy.

As a result, our calculations are limited only by the number of
radial grid points $N_r$ (we set $N_r=576$ for larger atoms),
the number of spherical harmonics included (we include up to $L=10$),
and the cutoff radius of the grid $r_m$
(typically $r_m\geq 24~a_0$ for larger atoms
- at this radius tiny e.g. $n(r_m)\ll 10^{-12}$).
We thus expect our calculations to be very well converged, with
estimated numerical errors of no more than 1\% (as an absolute
worst case - we suspect most atoms and ions will be well within
0.5\% of their converged values). For example, we have tested
Lutetium ($Z=N=71$ - a likely ``bad'' case due to its open $p$-shell
and large number of electrons) against increases
in the abscissae number and grid radius and found variation
of under 0.2\% in the $C_6$ coefficient and static polarizability
(e.g. the $C_6$ coefficient ranges
from 2546.2 for $N_r=576$ and $r_m=30.0$
to 2547.1 for $N_r=640$ and $r_m=24.0$,
and 2547.7 for $N_r=640$ and $r_m=30.0$).

\subsection{Corrected TDDFT calculations}
\label{sec:Correction}

While the TDDFT calculations described above
give moderately accurate polarizabilities (often within 10\% of
high level theories) and $C_6$ coefficients (often within 20\%),
they cannot meet quantum chemical accuracy.
To obtain true benchmarks we thus need to correct the TDDFT
results using higher level calculations or experimental data.

Let us fist consider the main sources of fundamental,
methodological errors in our approach (as opposed to numerical
errors, which we estimate to be under 1\%,
see Sec.~\ref{sec:converge}).
In error free calculations, the real frequency poles of
$\chi$ and $\chi_0$ are respectively the excitation energies
$E_k-E_l$ (for dipole transition excitations only),
and the excitation energies of the Kohn-Sham system
$\epsilon_k-\epsilon_l$ (again for dipole transitions only).
Furthermore, the set of full transitions $E_k-E_l$ is related
to the set of Kohn-Sham, transitions $\epsilon_k-\epsilon_l$
with corrections dependending on $f_{\Hxc}$. This relationship
can be approximated as $E_k-E_l\approx \epsilon_k-\epsilon_l
+ \ibraketop{kl}{f_{\Hxc}}{kl}$.
Methodological errors thus appear in two ways:
i) via the Kohn-Sham orbitals $\phi_i$ and their
energies $\epsilon_i$ which
are inaccurate due to errors in $v_s$
[see Eqs~\eqref{eqn:chi0} and \eqref{eqn:HamG}];
and ii) via the approximate kernels which introduces
additional errors to $\chi$ via the screening equation
\eqref{eqn:chi}.

These errors influence the quality of the polarizability
mostly via the lowest energy transition, a fact that we
will exploit. Using perturbation theory we can write
\begin{align}
  \alpha(i\omega)\approx& \sum_{kl}
  \frac{d_{kl}}{(E_k-E_l)^2+\omega^2}
  \label{eqn:alphaFormal}
\end{align}
where $\sum d_{kl}=N$.
Setting $\omega=0$ in \eqref{eqn:alphaFormal} shows that
the static polarizability $\alpha(0)$ is dominated by the
lowest energy transitions provided the dipole factor $d_{kl}$
is sufficiently large (which has been confirmed for
many atoms and ions\cite{Mitroy2010}).
$\alpha(0)$ is consequently more susceptible to errors in the
energy difference than $\alpha(i\omega>0)$.
We thus expect that sensibly correcting for errors in $\alpha(0)$
will go a long way to correcting errors in $\alpha(i\omega)$
for all $\omega$.

Following Chu and Dalgarno\cite{Chu2004} we perform
frequency rescaling to improve our frequency dependent
polarizabilities and $C_6$ coefficients. We replace the
raw TDDFT polarizability $\alpha^{\text{TDDFT}}$ by
\begin{align}
  \alpha(i\omega)\approx& S^2\alpha^{\text{TDDFT}}(iS\omega)
  \label{eqn:rescaling}
\end{align}
where $S=\sqrt{\alpha^{\text{Ref.}}(0)/\alpha^{\text{TDDFT}}(0)}$
is our scaling factor.
This form guarantees $\alpha(i\omega\to\infty)\to N/\omega^2$
and also ensures that
\begin{align}
  \alpha(0)=&\alpha^{\text{Ref.}}(0).
\end{align}
where $\alpha^{\text{Ref.}}$ is a reference benchmark static
polarizability. Under the assumption that the low frequency
polarizability is dominated by the smallest transition
energy this is equivalent to correcting a poor TDDFT
lowest transition energy using higher level data.

For our benchmarks we utilize a variety of different sources
(\rcites{Sadlej1991,Miadokova1997,Holka2005,Schwerdtfeger2006,Thierfelder2008,Dzuba2014,Jiang2015,Ehn2016})
to find static polarizabilities that we consider to be
optimal reference values. These are discussed in Section~1
of the Supplementary Material\cite{Supp}.
For most ions and Row 6 elements accurate dipole data are not
available. In cases where data are not available we use:
a) for neutral atoms we use an average scaling factor obtained
from known cases in the same row;
b) for ions we use the scaling coefficient of the atom
with the same number of electrons.

We also produce a second set of data based on rescaled TDDFT
with the PGG kernel that utilises a ``minimal chemistry'' model
described in Appendix~\ref{app:MinChem}. This allows us to
extend our results to double anions and give a (hopefully)
more realistic depiction of embedded open shell single anions while
maintaining the presumed good results for atoms and cations.
These polarizabilities are not intended to be
compared to experimental data. But they provide a potentially
more realistic starting point for atom-in-molecule approaches
like vdW functionals and semi-classical molecular modelling.

\section{Benchmarking tests}

\begin{table}
  \caption{Comparison of our scaled $C_6$ values with
    high-level wave-function method reference values.
    Results in Ha$a_0^6$.
    \label{tab:C6CompareBenchmark}}
  %\begin{ruledtabular}%
    \begin{tabular}{lrrrrl}
      Species & Ref. value & This work & Err \% & Ref.
      \\\hline
 He     & 1.46 & 1.47 & 0.9 & \cite{Jiang2015} \\
 Li     & 1396 & 1408 & 0.9 & \cite{Jiang2015} \\
 Be$^+$ & 68.8 & 70.3 & 2.2 & \cite{Jiang2015} \\
 Be  & 213 & 214 & 0.3 & \cite{Jiang2015} \\
 Ne & 6.38 & 6.91 & 8.4 & \cite{Jiang2015} \\
 Na & 1562 & 1566 & 0.2 & \cite{Jiang2015} \\
 Mg$^+$ & 155 & 155 & 0.5 & \cite{Jiang2015} \\
 Mg & 630 & 629 & -0.1 & \cite{Jiang2015} \\
 Ar & 64.3 & 67.4 & 4.8 & \cite{Jiang2015} \\
 K  & 3906 & 3914 & 0.2 & \cite{Jiang2015} \\
 Ca$^+$ & 541 & 554 & 2.3 & \cite{Jiang2015} \\
 Ca & 2188 & 2232 & 2.0 & \cite{Jiang2015} \\
 Cu & 250 & 264 & 5.5 & \cite{Jiang2015} \\
 Kr & 130 & 136 & 4.8 & \cite{Jiang2015} \\
 Rb & 4667 & 4660 & -0.1 & \cite{Jiang2015} \\
 Sr$^+$ & 776 & 790 & 1.9 & \cite{Jiang2015} \\
 Sr & 3149 & 3230 & 2.6 & \cite{Jiang2015} \\
 Ag & 342 & 341 & -0.3 & \cite{Jiang2015} \\
 Xe & 286 & 302 & 5.6 & \cite{Jiang2015} \\
 Cs & 6733 & 6657 & -1.1 & \cite{Jiang2015} \\
 Ba$^+$ & 1293 & 1296 & 0.3 & \cite{Jiang2015} \\
 Ba & 5380 & 5543 & 3.0 & \cite{Jiang2015} \\
\hline
MAE & & & & 2.2 \\
ME  & & & & 2.0 \\
    \end{tabular}%
  %\end{ruledtabular}
  %\\$\ddag$ Average over listed theory values
\end{table}
In our first test we check the accuracy of the rescaling
of TDDFT results using equation~\eqref{eqn:rescaling}.
In Table~\ref{tab:C6CompareBenchmark} we compare $C_6$
coefficients from this approach with the same $C_6$ coefficients
calculated using accurate quantum chemical theories.
As can be seen, the agreement is generally
very good, with the worst case (Ne) off by 8.4\%
(we cannot explain this difference, but note that both
PGG and RXH kernels give similar results),
and the second worst case (Cr) only 5.5\% off
the reference results despite our use of a different
symmetry state. The $C_6$ coefficients obtained from
our approach are also similar to those found by
Chu and Dalgarno\cite{Chu2004}.

\begin{figure}
  \caption{Normalised RXH $C_6$ coefficients versus PGG
    values. The dotted lines indicate a difference
    of 20\% between the two methods. Note that the worst
    outliers are all elements for which the rescaling parameter
    was interpolated based on neighbouring elements.
    \label{fig:CompareFunctionals}}
  \includegraphics[width=\linewidth,clip]{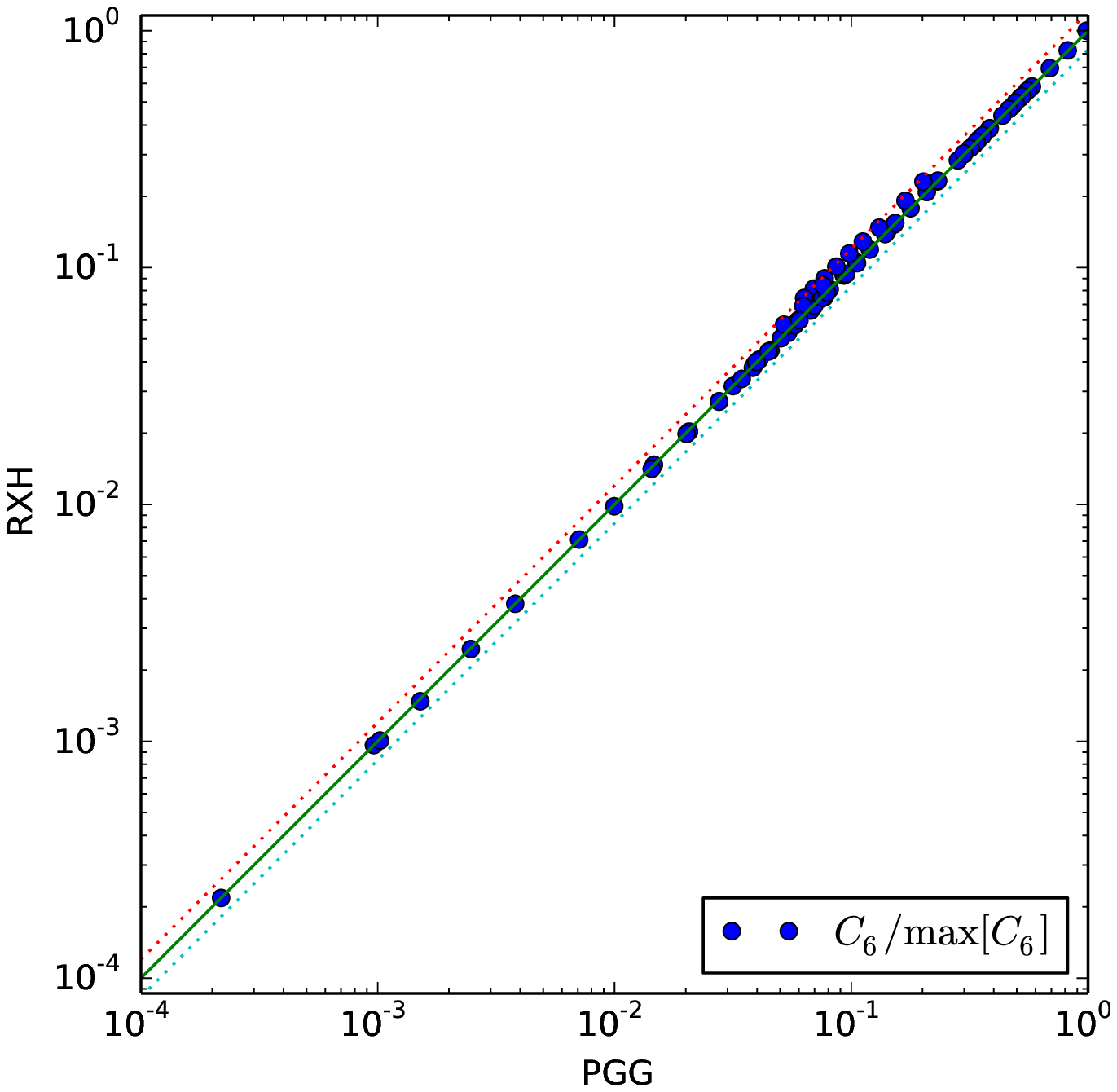}
\end{figure}
By using different TDDFT kernels we are able to
provide a second test of the rescaling that incldues systems
for which quantum chemical data are unavailable.
Our test is based on the following idea:
if rescaling can give good dynamic polarizabilities by
correcting TDDFT calculations using
\emph{static} polarizabilities, then rescaled $C_6$ coefficients
of different species should be largely independent of the
exchange-correlation kernel employed in the TDDFT calculations.
To test this hypothesis,
we carried out calculations using two kernels:
that of Petersilka-Gossman and Gross\cite{PGG}, and
the ``radial exchange hole'' kernel of Gould\cite{Gould2012-RXH}.
These two approaches give very different dipole polarizabilities
before scaling, and can thus provide a clear test of the
rescaling.

The results of our tests are plotted in
Figure~\ref{fig:CompareFunctionals}, in which rescaled RXH atomic
$C_6$ coefficents are shown as a function of their PGG values.
If all points fell perfectly on the straight line this would
indicate that the rescaling was perfect, as two sets of inputs
would have yielded the same set of $C_6$ coefficients after scaling.
As it is, the values are all within 20\% of each other,
with the worst cases \emph{all} being elements for which accurate
reference polarizability values
were not available and the rescaling parameters had to be interpolated
based on other elements in the same row. 

We thus conclude that starting with a moderately accurate TDDFT
kernel, and then rescaling based on static dipole polarizabilities,
is a very accurate approach for determining dynamic polarizabilities.
Not only do we get good agreement with more accurate approaches,
but we also get agreement across three TDDFT approaches:
that of Chu and Dalgarno\cite{Chu2004}, and the two kernels
tested here.

\section{Results}

\begin{table*}[t]
  \caption{Static polarizabilities and $C_6$ coefficients
    from the ``benchmark'' data set for all neutral atoms.
    We note that the small error in Hydrogen's $C_6$ coefficient
    of 13/2 is an artefact of the two-Lorentzian model.
    Results in $a_0^3$ for polarizabilities and Ha$a_0^6$ for
    $C_6$ coefficients.
    \label{tab:C6-0}}
  {\tiny
  %\begin{ruledtabular}%
  \begin{tabular}{lrrlrrlrrlrrlrrlrr}
    ID & $\alpha(0)$ & $C_6$ &
    ID & $\alpha(0)$ & $C_6$ &
    ID & $\alpha(0)$ & $C_6$ &
    ID & $\alpha(0)$ & $C_6$ &
    ID & $\alpha(0)$ & $C_6$ &
    ID & $\alpha(0)$ & $C_6$ \\
    \hline
         H &   4.50 &   6.51&
        He &   1.38 &   1.47\\

        Li &    164 &   1410&
        Be &   37.7 &    214\\

         B &   20.5 &   99.2&
         C &   11.7 &   47.9&
         N &   7.25 &   25.7&
         O &   5.20 &   16.7&
         F &   3.60 &   10.2&
        Ne &   2.67 &   6.91\\

        Na &    163 &   1570&
        Mg &   71.4 &    629\\

        Al &   57.5 &    520&
        Si &   37.0 &    308&
         P &   24.8 &    187&
         S &   19.5 &    140&
        Cl &   14.7 &   97.1&
        Ar &   11.1 &   67.4\\

         K &    290 &   3910&
        Ca &    160 &   2230\\

        Sc &    123 &   1570&
        Ti &    102 &   1200&
         V &   87.3 &    955&
        Cr &   78.4 &    709&
        Mn &   66.8 &    635&
        Fe &   60.4 &    548\\
        Co &   53.9 &    461&
        Ni &   48.4 &    393&
        Cu &   41.7 &    264&
        Zn &   38.4 &    276\\

        Ga &   52.1 &    456&
        Ge &   40.2 &    365&
        As &   29.6 &    260&
        Se &   26.2 &    233&
        Br &   21.6 &    187&
        Kr &   16.8 &    136\\

        Rb &    317 &   4660&
        Sr &    198 &   3230\\

         Y &    163 &   2600&
        Zr &    112 &   1360&
        Nb &   97.9 &   1140&
        Mo &   87.1 &   1030&
        Tc &   79.6 &    939&
        Ru &   72.3 &    809\\
        Rh &   66.4 &    708&
        Pd &   61.7 &    628&
        Ag &   46.2 &    341&
        Cd &   46.7 &    405\\

        In &   62.1 &    643&
        Sn &   60.0 &    715&
        Sb &   44.0 &    504&
        Te &   40.0 &    471&
         I &   33.6 &    389&
        Xe &   27.2 &    302\\

        Cs &    396 &   6660&
        Ba &    278 &   5540&
        La &    214 &   3730&
        Ce &    205 &   3480&
        Pr &    216 &   3760&
        Nd &    209 &   3560\\
        Pm &    200 &   3340&
        Sm &    192 &   3130&
        Eu &    184 &   2940&
        Gd &    158 &   2340&
        Tb &    170 &   2590&
        Dy &    163 &   2430\\
        Ho &    156 &   2280&
        Er &    150 &   2150&
        Tm &    144 &   2020&
        Yb &    139 &   1910\\

        Lu &    137 &   2020&
        Hf &   83.7 &   1040&
        Ta &   73.9 &    887&
         W &   65.8 &    757&
        Re &   60.2 &    663&
        Os &   55.3 &    584\\
        Ir &   51.3 &    522&
        Pt &   48.0 &    470&
        Au &   45.4 &    427&
        Hg &   33.5 &    268\\

        Tl &   51.4 &    509&
        Pb &   47.9 &    534&
        Bi &   43.2 &    513&
        Po &   36.1 &    424&
        At &   30.4 &    351&
        Rn &   32.2 &    408\\

  \end{tabular}%
  %\end{ruledtabular}
  }
\end{table*}
\begin{table*}[t]
  \caption{Static polarizabilities and $C_6$ coefficients
    from the ``benchmark'' data set for all cations.
    Results in $a_0^3$ for polarizabilities and Ha$a_0^6$ for
    $C_6$ coefficients.
    \label{tab:C6--1}}
  {\tiny
  %\begin{ruledtabular}%
  \begin{tabular}{lrrlrrlrrlrrlrrlrr}
    ID & $\alpha(0)$ & $C_6$ &
    ID & $\alpha(0)$ & $C_6$ &
    ID & $\alpha(0)$ & $C_6$ &
    ID & $\alpha(0)$ & $C_6$ &
    ID & $\alpha(0)$ & $C_6$ &
    ID & $\alpha(0)$ & $C_6$ \\
    \hline
  He$^{+}$ &  0.294 &  0.109&
  Li$^{+}$ &  0.193 &  0.079\\

  Be$^{+}$ &   24.5 &   70.3&
   B$^{+}$ &   9.67 &   25.2\\

   C$^{+}$ &   5.66 &   13.3&
   N$^{+}$ &   3.68 &   8.02&
   O$^{+}$ &   2.55 &   5.27&
   F$^{+}$ &   1.78 &   3.44&
  Ne$^{+}$ &   1.44 &   2.75&
  Na$^{+}$ &  0.930 &   1.54\\

  Mg$^{+}$ &   35.0 &    155&
  Al$^{+}$ &   19.6 &   89.7\\

  Si$^{+}$ &   18.2 &   94.8&
   P$^{+}$ &   14.3 &   74.1&
   S$^{+}$ &   11.9 &   62.2&
  Cl$^{+}$ &   9.27 &   46.1&
  Ar$^{+}$ &   7.28 &   34.2&
   K$^{+}$ &   5.05 &   21.0\\

  Ca$^{+}$ &   75.5 &    554&
  Sc$^{+}$ &   60.0 &    531\\

  Ti$^{+}$ &   44.6 &    286&
   V$^{+}$ &   36.5 &    217&
  Cr$^{+}$ &   38.2 &    236&
  Mn$^{+}$ &   25.8 &    132&
  Fe$^{+}$ &   23.2 &    114&
  Co$^{+}$ &   20.6 &   95.8\\
  Ni$^{+}$ &   18.4 &   81.5&
  Cu$^{+}$ &   17.6 &   75.9&
  Zn$^{+}$ &   17.9 &   91.7&
  Ga$^{+}$ &   15.2 &   71.8\\

  Ge$^{+}$ &   18.5 &    108&
  As$^{+}$ &   16.4 &    102&
  Se$^{+}$ &   15.7 &    104&
  Br$^{+}$ &   13.6 &   91.2&
  Kr$^{+}$ &   11.1 &   71.9&
  Rb$^{+}$ &   8.32 &   49.1\\

  Sr$^{+}$ &   90.2 &    790&
   Y$^{+}$ &   88.1 &   1020\\

  Zr$^{+}$ &   62.0 &    574&
  Nb$^{+}$ &   55.3 &    512&
  Mo$^{+}$ &   47.6 &    424&
  Tc$^{+}$ &   32.7 &    246&
  Ru$^{+}$ &   37.9 &    327&
  Rh$^{+}$ &   34.4 &    284\\
  Pd$^{+}$ &   31.5 &    248&
  Ag$^{+}$ &   20.0 &    115&
  Cd$^{+}$ &   23.1 &    155&
  In$^{+}$ &   20.2 &    127\\

  Sn$^{+}$ &   29.2 &    237&
  Sb$^{+}$ &   25.5 &    215&
  Te$^{+}$ &   25.1 &    230&
   I$^{+}$ &   22.1 &    205&
  Xe$^{+}$ &   18.7 &    170&
  Cs$^{+}$ &   15.0 &    129\\

  Ba$^{+}$ &    121 &   1300&
  La$^{+}$ &   94.2 &   1180&
  Ce$^{+}$ &   89.1 &   1080&
  Pr$^{+}$ &   93.4 &   1160&
  Nd$^{+}$ &   89.6 &   1080&
  Pm$^{+}$ &   85.7 &   1010\\
  Sm$^{+}$ &   81.8 &    941&
  Eu$^{+}$ &   78.1 &    876&
  Gd$^{+}$ &   67.2 &    698&
  Tb$^{+}$ &   71.8 &    769&
  Dy$^{+}$ &   68.6 &    717&
  Ho$^{+}$ &   65.8 &    672\\
  Er$^{+}$ &   63.0 &    630&
  Tm$^{+}$ &   60.4 &    590&
  Yb$^{+}$ &   58.0 &    555&
  Lu$^{+}$ &   74.6 &    808\\

  Hf$^{+}$ &   47.5 &    443&
  Ta$^{+}$ &   42.0 &    388&
   W$^{+}$ &   37.1 &    333&
  Re$^{+}$ &   33.4 &    290&
  Os$^{+}$ &   30.4 &    256&
  Ir$^{+}$ &   27.9 &    228\\
  Pt$^{+}$ &   25.8 &    204&
  Au$^{+}$ &   24.0 &    184&
  Hg$^{+}$ &   17.5 &    114&
  Tl$^{+}$ &   17.0 &    109\\

  Pb$^{+}$ &   23.5 &    185&
  Bi$^{+}$ &   25.2 &    226&
  Po$^{+}$ &   22.9 &    213&
  At$^{+}$ &   20.2 &    190&
  Rn$^{+}$ &   22.4 &    236&
  \end{tabular}%
  %\end{ruledtabular}
  }
\end{table*}
\begin{table*}[t]
  \caption{Static polarizabilities and $C_6$ coefficients
    from the ``benchmark'' data set for all anions.
    Results in $a_0^3$ for polarizabilities and Ha$a_0^6$ for
    $C_6$ coefficients.
    \label{tab:C6-1}}
  {\tiny
  %\begin{ruledtabular}%
  \begin{tabular}{lrrlrrlrrlrrlrrlrr}
    ID & $\alpha(0)$ & $C_6$ &
    ID & $\alpha(0)$ & $C_6$ &
    ID & $\alpha(0)$ & $C_6$ &
    ID & $\alpha(0)$ & $C_6$ &
    ID & $\alpha(0)$ & $C_6$ &
    ID & $\alpha(0)$ & $C_6$ \\
    \hline
   H$^{-}$ &    216 &   2400\\

  Li$^{-}$ &   1180 &  35170\\

   B$^{-}$ &   32.9 &    232&
   C$^{-}$ &   15.5 &   81.7&
   N$^{-}$ &   8.04 &   33.1&
   O$^{-}$ &   5.40 &   19.1&
   F$^{-}$ &   15.0 &   73.5\\

  Na$^{-}$ &   1310 &  42690\\

  Al$^{-}$ &    109 &   1550&
  Si$^{-}$ &   59.6 &    698&
   P$^{-}$ &   33.7 &    322&
   S$^{-}$ &   24.1 &    206&
  Cl$^{-}$ &   30.3 &    276\\

   K$^{-}$ &   2090 &  87490\\

  Sc$^{-}$ &    134 &   1860&
  Ti$^{-}$ &    110 &   1380&
   V$^{-}$ &   94.5 &   1100&
  Cr$^{-}$ &   83.5 &    908&
  Mn$^{-}$ &   74.7 &    762&
  Fe$^{-}$ &   66.5 &    640\\
  Co$^{-}$ &   59.8 &    545&
  Ni$^{-}$ &   54.2 &    470&
  Cu$^{-}$ &    500 &   9440\\

  Zn$^{-}$ &    845 &  17790&
  Ga$^{-}$ &    103 &   1430&
  Ge$^{-}$ &   63.3 &    796&
  As$^{-}$ &   40.0 &    443&
  Se$^{-}$ &   31.7 &    332&
  Br$^{-}$ &   42.8 &    497\\

  Rb$^{-}$ &   2110 &  92240\\

   Y$^{-}$ &    175 &   3080&
  Zr$^{-}$ &    134 &   2160&
  Nb$^{-}$ &    114 &   1680&
  Mo$^{-}$ &    100 &   1390&
  Tc$^{-}$ &   90.6 &   1170&
  Ru$^{-}$ &   82.4 &   1010\\
  Rh$^{-}$ &   75.9 &    881&
  Pd$^{-}$ &   70.5 &    779&
  Ag$^{-}$ &    501 &   9800\\

  Cd$^{-}$ &    863 &  20230&
  In$^{-}$ &    133 &   2240&
  Sn$^{-}$ &   89.1 &   1410&
  Sb$^{-}$ &   59.6 &    859&
  Te$^{-}$ &   49.1 &    684&
   I$^{-}$ &   61.7 &    925\\

  Cs$^{-}$ &   2480 & 118890&
  La$^{-}$ &    729 &  19310&
  Ce$^{-}$ &   4.02 &   46.5&
  Pr$^{-}$ &   4.88 &   62.7&
  Nd$^{-}$ &   5.57 &   77.0&
  Pm$^{-}$ &   6.50 &   97.9\\
  Sm$^{-}$ &   7.38 &    119&
  Eu$^{-}$ &   8.08 &    138&
  Gd$^{-}$ &   8.08 &    139&
  Tb$^{-}$ &   9.84 &    188&
  Dy$^{-}$ &   10.5 &    210&
  Ho$^{-}$ &   11.1 &    230\\
  Er$^{-}$ &   11.6 &    246&
  Tm$^{-}$ &   12.0 &    260\\

  Lu$^{-}$ &   13.8 &    325&
  Hf$^{-}$ &    236 &   4170&
  Ta$^{-}$ &    206 &   3640&
   W$^{-}$ &    295 &   5350&
  Re$^{-}$ &    368 &   6710&
  Os$^{-}$ &    378 &   6840\\
  Ir$^{-}$ &    381 &   6830&
  Pt$^{-}$ &    383 &   6810&
  Au$^{-}$ &    387 &   6840\\

  Tl$^{-}$ &    356 &   6820&
  Pb$^{-}$ &    135 &   2310&
  Bi$^{-}$ &    103 &   1710&
  Po$^{-}$ &   73.1 &   1140&
  At$^{-}$ &   54.5 &    801\\

  \end{tabular}%
  %\end{ruledtabular}
  }
\end{table*}
We show in Tables~\ref{tab:C6-0}-\ref{tab:C6-1}
static polarizabilities and same-species
$C_6$ coefficients using the ``benchmark'' data set
of rescaled atoms and ions. These are arranged by shell
structure to make for easy access. As discussed previously,
these are in good agreement with external reference data
and, for the most part, are in good agreement with RXH
values. The exceptions show no more than a 15\%
difference, suggesting that this is a good estimate for
the worst-case accuracy of our approach.

In addition to the tabulated static polarizabilities and
$C_6$ coefficients, we also provide
readily usable data for all cases considered here.
Rather than provide tabulated frequency dependent polarizability
data for all 411 atoms and ions (or 443 for the minimal chemistry
model), we instead provide parameters $a_{c,X}$ and $\Omega_{c,X}$
where $c=1,2$ for a two Lorentzian model
\begin{align}
  \alpha_X(i\omega)\approx&
  \sum_{c=1,2} \frac{a_{c,X}}{\omega^2 + \Omega_{c,X}^2},
  \label{eqn:alphaModel}
\end{align}
for species $X$ with nuclear charge $Z$ and $N$ electrons.
Eq.~\eqref{eqn:alphaModel}
can then be used to accurately reproduce the polarizabilities
for arbitrary frequency.

This two-Lorentzian representation of
polarizabilities can be used to reproduce all $C_6$ coefficients
using analytic solutions of the Casimir-Polder formula 
\eqref{eqn:CP} which take the form (here $X\equiv Z_X,N_X$
and $Y\equiv Z_YN_Y$)
\begin{align}
  C_{6,XY}=&\sum_{cc'}
  \frac{3a_{c,X}a_{c',Y}}%
       {2\Omega_{c,X}\Omega_{c',Y}
         [\Omega_{c,X}+\Omega_{c',Y}]}.
       \label{eqn:twolor}
\end{align}
Results are within a few percent compared of similar calculations
carried out on the raw data
(see Appendix~\ref{app:TwoLor} for details). Our two-Lorentzian model
thus provides a useful and compact representation for
all dynamic polarizabilities, allowing reconstruction of
the 84000+ different-species $C_6$ coefficients.
These data are included in the supplementary material\cite{Supp},
tabulated for the ``benchmark'' set and included in ascii
files for all data sets.

\begin{table}
  \caption{Interaction of rare gas atoms with hailde atoms
    and ions and Na$^+$.
    Results in Ha$a_0^6$.
    \label{tab:RG}}
  \begin{tabular}{lrrrl}
    & $C _6$ [Ref] & $C_6$ [this work] & \% Err & Ref.
    \\\hline
  Br$^+$--He &   12.0 &   11.3 &  -5.8 & \cite{Buchachenko2009} \\
      Br--He &   15.0 &   15.8 &   5.1 & \cite{Buchachenko2009} \\
  Br$^-$--He &   27.0 &   24.0 & -11.0 & \cite{Buchachenko2009} \\
  Br$^+$--Ne &   23.0 &   23.7 &   3.2 & \cite{Buchachenko2009} \\
      Br--Ne &   31.0 &   32.8 &   5.8 & \cite{Buchachenko2009} \\
  Br$^-$--Ne &   48.0 &   49.4 &   2.9 & \cite{Buchachenko2009} \\
  Br$^+$--Ar &   78.0 &   78.3 &   0.4 & \cite{Buchachenko2009} \\
      Br--Ar &    110 &    111 &   0.9 & \cite{Buchachenko2009} \\
  Br$^-$--Ar &    174 &    174 &  -0.3 & \cite{Buchachenko2009} \\
\hline
  Na$^+$--He &   1.79 &   1.39 & -22.2 & \cite{Soldan1999} \\
  Na$^+$--Ne &   3.22 &   3.15 &  -2.2 & \cite{Soldan1999} \\
  Na$^+$--Ar &   10.4 &   8.71 & -15.9 & \cite{Soldan1999} \\
\hline
   F$^-$--He &   9.37 &   9.81 &   4.7 & \cite{Hattig1998} \\
   F$^-$--Ne &   19.4 &   20.5 &   5.6 & \cite{Hattig1998} \\
   F$^-$--Ar &   66.4 &   69.0 &   3.8 & \cite{Hattig1998} \\
   F$^-$--Kr &   95.2 &   98.9 &   3.9 & \cite{Hattig1998} \\
   F$^-$--Xe &    143 &    148 &   3.7 & \cite{Hattig1998} \\
   F$^-$--Rn &    170 &    172 &   1.4 & \cite{Hattig1998} \\
  Cl$^-$--He &   19.1 &   18.3 &  -4.0 & \cite{Hattig1998} \\
  Cl$^-$--Ne &   39.4 &   37.8 &  -4.1 & \cite{Hattig1998} \\
  Cl$^-$--Ar &    138 &    131 &  -4.8 & \cite{Hattig1998} \\
  Cl$^-$--Kr &    198 &    190 &  -4.3 & \cite{Hattig1998} \\
  Cl$^-$--Xe &    299 &    286 &  -4.3 & \cite{Hattig1998} \\
  Cl$^-$--Rn &    357 &    333 &  -6.7 & \cite{Hattig1998} \\
  Br$^-$--He &   24.0 &   24.0 &   0.2 & \cite{Hattig1998} \\
  Br$^-$--Ne &   49.4 &   49.4 &  -0.0 & \cite{Hattig1998} \\
  Br$^-$--Ar &    174 &    174 &  -0.3 & \cite{Hattig1998} \\
  Br$^-$--Kr &    251 &    251 &   0.0 & \cite{Hattig1998} \\
  Br$^-$--Xe &    380 &    380 &   0.1 & \cite{Hattig1998} \\
  Br$^-$--Rn &    452 &    443 &  -1.9 & \cite{Hattig1998} \\
   I$^-$--He &   32.1 &   32.1 &  -0.0 & \cite{Hattig1998} \\
   I$^-$--Ne &   65.9 &   65.8 &  -0.2 & \cite{Hattig1998} \\
   I$^-$--Ar &    233 &    233 &  -0.0 & \cite{Hattig1998} \\
   I$^-$--Kr &    336 &    338 &   0.5 & \cite{Hattig1998} \\
   I$^-$--Xe &    510 &    514 &   0.7 & \cite{Hattig1998} \\
   I$^-$--Rn &    608 &    599 &  -1.5 & \cite{Hattig1998} \\
  \end{tabular}
\end{table}
We also test interactions of rare gas atoms with halides
and the alkali cation  Na$^{+}$, to demonstrate the method's
capabilities on heteronuclear $C_6$ coefficients.
We compare results from our data set against results from
three references, and are thus able to cover elements
in all of the six rows considered here. Results are
presented in Table~\ref{tab:RG}. Agreement is generally
excellent, with a few cases substantially worse than
the mean absolute relative error of 3.7\%.

The worst relative errors all involve Na$^+$ and rare gases.
However, the absolute errors for these cases are small
since neither Na$^+$ nor rare gas atoms are very polarizable
and the $C_6$ coefficients are consequently small.
Also, the data from \rcite{Soldan1999} may not be as accurate
as more modern calculations would allow. The next worst
case is Br$^-$ with He when compared with the reference
data of \rcite{Buchachenko2009}. Oddly, we get almost
perfect agreement for this case with the
(presumably less accurate) TDMP2 results
of \rcite{Hattig1998}.

\begin{figure}
  \caption{Dispersion coefficients for pair of species $X$ and $Y$
      (neutral atoms from first six rows of PT) plotted against
    $\alpha_{XY}(0)\equiv \sqrt{\alpha_X(0)\alpha_Y(0)}$
    (red crosses) compared with best fit function (blue line).
    Units are $a_0^3$ for polarizabilities and Ha$a_0^6$ for
    $C_6$ coefficients.
    Inset shows the same as the main plot but zooms into the most
    populated range of polarizabilities and coefficients
    \label{fig:C6A0}}
  \includegraphics[width=\linewidth,clip]{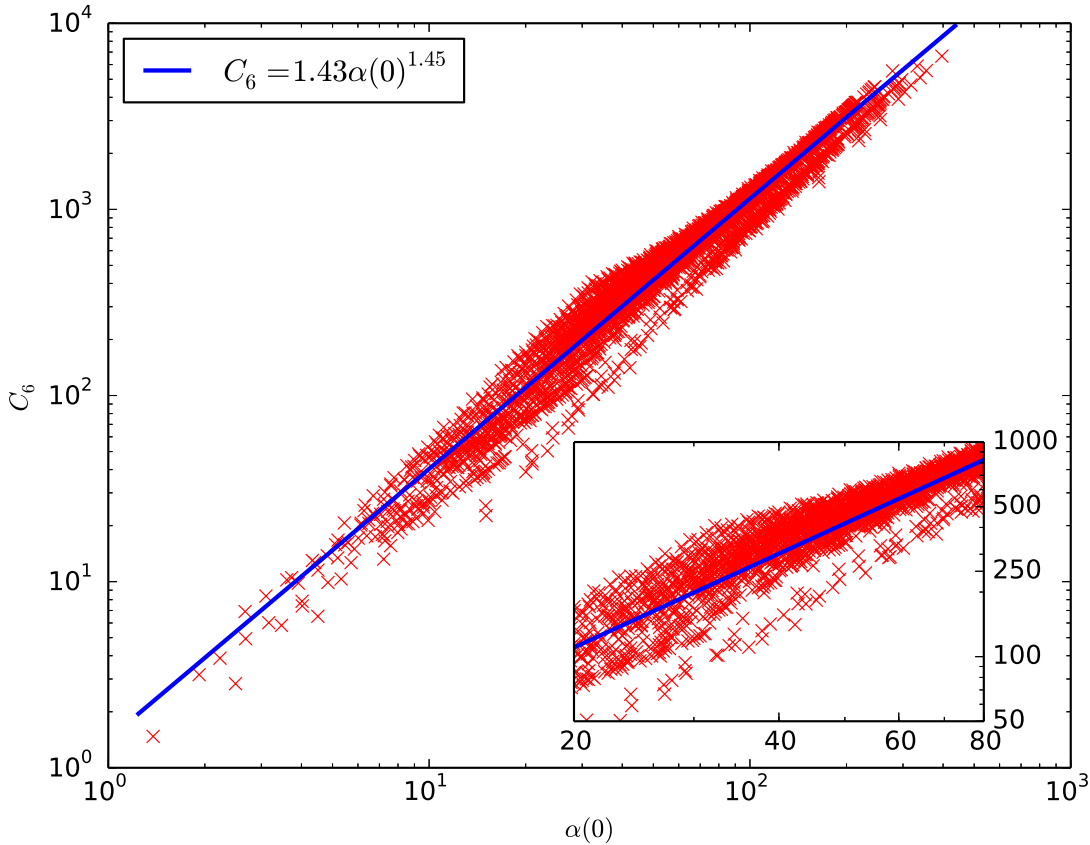}
\end{figure}
Finally, we use our comprehenesive data set to test for broad
relationships between $C_6$ coefficients and static dipole
polarizabilities $\alpha(0)$. Based on units, one predicts
that $C_{6,XY}\propto \alpha_X(0)\alpha_Y(0)$ might be a reasonable
approximation for such a relationship. However, after testing
all 3741 possible pairs of neutral atoms
(plotted in Figure~\ref{fig:C6A0}) we find that
\begin{align}
  C_{6,XY}\approx \Xi
  \big[\alpha_X(0)\alpha_Y(0)\big]^{0.73\pm 0.01},
\end{align}
$\Xi=(1.5\pm 0.1)$~[Ha$a_0^{1.62}$],
is a better power law fit, showing a somewhat surprising reduction
in the scaling coefficient from the expected product form.
Here our parameter error bars are
crudely estimated by comparing more limited fits with only
same-species atomic and ionic coefficients, and
all neutral atom pairs using our PGG and RXH data sets.

The constant prefactor $\Xi$ varies
considerably depending on the data set selected,
and Figure~\ref{fig:C6A0}  makes it clear that the actual
results have a much wider range of values. By contrast, the
exponent $0.73\pm 0.01$ remains essentially constant. This suggests
a certain universality of the power law relationship between
polarizabilities and $C_6$ coefficients.
We suspect that it might be possible to improve the quality of this
fit by improving the dependence on $\alpha_x(0)$ and $\alpha_y(0)$
separately.

Along these lines, we test the quality of the relationship
\begin{align}
  C_{6,XY}=\frac{2C_{6,XX}C_{6,YY}}%
  {\frac{\alpha_X(0)}{\alpha_Y(0)}C_{6,YY}
    + \frac{\alpha_Y(0)}{\alpha_X(0)}C_{6,XX}},
  \label{eqn:onelor}
\end{align}
found\cite{Tang1969} from a one-Lorentzian model
$\alpha(i\omega)=\alpha(0)/(1+\omega^2/\eta^2)$
(equivalent to the [1,0] Pad\'e approximation for
the polarizability).
This has previously been used (e.g. \rcite{Tkatchenko2009})
to derive different-species coefficients from same-species data.
We find that \eqref{eqn:onelor} is generally very accurate,
giving answers within 5\% of our two-Lorentzian model
\eqref{eqn:twolor} for almost 80\% of cases (from 84000 pairs
of atoms, anions and cations).
However in 2.0\% ($\approx 1700$) of cases it
is more than 20\% out
and in 0.2\% ($\approx 200$) of cases it
is more than 30\% out.

It should be noted that most of these worst case examples
involve 2+ or 3+ cations interacting with an anion, and
thus give rise to small \emph{absolute} errors. None of
the cases with $>20\%$ errors
involve two neutral atoms. Some notable bad cases
examples are Na$^+$ with Cs (-21\%), and
Al$^-$ with many anions in Row 6 ($\approx-24\%$). 

\section{Conclusions}

Using TDDFT with the PGG\cite{PGG} kernel
we calculated all-electron dipole polarizabilities of
all atoms and many cations and anions from rows 1-6 ($1\leq Z\leq 86$)
of the periodic table. We also performed calculations using
an alternative RXH\cite{Gould2012-RXH} kernel; and using a variant
method designed to approximate more realistic environmental
effects for atoms in molecules, which yields different
coefficients for open-shell anions.
We argue that these dipole polarizabilities can 
provide a rough benchmark (likely within 15\% for all species,
with better results expected in some cases)
for imaginary frequency dipole polarizability calculations
of $C_6$ coefficients.
They are almost certainly of sufficient quality to be used
in atom-in-molecule (AIM) approaches,
be they classical, semi-classical, or semi-empirical.

Our polarizabilities were parametrised using a two-Lorentzian model
\eqref{eqn:alphaModel} that can be used to reproduce $C_6$
coefficients within 5\% (and typically $<$1\%) of the value
obtained by full quadrature via the Casimir-Polder formula
\eqref{eqn:CP}. These parameters were tabulated for all
atoms and ions investigated, and are included in the
supplementary materials\cite{Supp}.
These were used to calculate homonuclear-isoelectronic
$C_6$ coefficients and some heteronuclear coefficients.

We finally used our data to study the dependence of $C_6$
coefficients on polarizabilities, finding
$C_{6,XY}\approx \Xi[\alpha_X(0)\alpha_Y(0)]^{0.73\pm 0.01}$,
with error bars indicating the spread of best-fit parameters
found on different data sets.
While the prefactor $\Xi$ was found to vary considerably depending
on the data set used to make the fit, the exponent
varied much less suggesting it is a more universal quantity.
Similarly, we tested the relationship
$C_{6,XY}=2C_{6,X}C_{6,Y}/[\alpha_X/\alpha_YC_{6,Y}+\alpha_Y/\alpha_XC_{6,X}]$
sometimes used to relate homonuclear-isoelectronic
coefficients and $C_6$ for pairs of unlike species (atoms or ions).
We found that it gave errors of less than 5\% in 80\% of
cases, but in 0.2\% of cases gave very poor results with
$>30$\% errors.

In future we also hope to use our approach to study atoms and ions
in Row 7, after developing techniques to deal with relativistic
effects and fixed core approximations. We also aim to explore other
environmental effects to better understand how embedded
atoms behave compared to their free counterparts, and thus
to improve and extend the ``minimal chemistry'' model
presently used to determine some cation and all double
cation dipole polarizabilities.

\begin{acknowledgement}
TG recognises computing support from the Griffith University Gowonda
HPC Cluster.
TB acknowledges support from the project APVV-15-0105 and
the use of computational
resources of supercomputing infrastructure of Computing
Center of the Slovak Academy of Sciences acquired in Project
Nos. ITMS 26230120002 and 26210120002 supported by the
Research and Development Operational Program funded by
the ERDF.
The authors thank Prof. Ivan {\v{C}}ernu{\v{s}}{\'a}k
for sharing accurate polarizability data for open-shell atoms
and ions.
\end{acknowledgement}

\begin{suppinfo}
We include with this manuscript supporting information including:
\begin{itemize}
\item[] A PDF document that includes: a) further details of the
benchmark data set; b) several additional tables
of chief results; c) a description of the accompanying ascii files;
\item[] Several ascii files presenting chief results in a
readily usable format.
\end{itemize}
\end{suppinfo}

\appendix

\section{Transition metal and lanthanide atoms and ions}
\label{app:TM}

Some of the transition metal and lanthanide atoms and ions represent
very difficult casees for single-determinant theories like DFT
and HF. In certain cases the $s$ and $d$ orbitals (or $d$
and $f$ orbitals) are so close
to degenerate that the aufbau principle can be violated in
the lowest \emph{energy} configuration for DFT calculations
(see e.g. \rcite{Johnson2007} or
\rcite{Koga1995} for discussion in a Hartree-Fock context).
Despite the LEXX approximation including
some multi-determinental characteristics via the ensemble
DFT formalism it too is affected by these issues. Furthermore,
the route of explicit symmetry breaking employed by
Johnson \latin{et al} is not available in our radial code.

This poses a particular challenge for polarizability problems
as the effective Kohn-Sham gap $\epsilon_l-\epsilon_h$
(for lowest unoccupied $f_l=0$ and highest occupied $f_h>0$
KS orbitals)
can become negative. Since this gap appears in the denominator
of terms contributing to $\chi_0$ it leads to unphysical small
frequency (real or imaginary) polarizabilities. In certain
cases it gives negative bare polarizabilities (i.e. calcualated
via \eqref{eqn:alpha} except with $\chi_0$ rather than $\chi$).
While these problems are somewhat mitigated in the interacting
response $\chi$ by the screening kernel $f_{\Hxc}$
(e.g. our approach returns positive static polarizabilities for all
species tested), it still contributes substantial errors to
polarizabilities.

We are thus left with a dilemma: do we choose the lowest
\emph{energy} groundstate or choose a state that does not
give negative transitions but is in a less realistic
electronic configuration? We thus decide to simplify matters
by filling the orbitals according to $(n+1)s^2(n)d^m$,
where $n=3$ or 4 (or equivalent for $d$ and $f$ orbitals).
This leads to some differences from free atom calculations,
especially in Cr and Mo which have clearly lower
energies in their $(n+1)s^1(n)d^5$ states. However, as these
elements usually appear as embedded ions we expect the
practical effect to be minimal.

\section{Minimal chemistry model}
\label{app:MinChem}

The main purpose of this manuscript is to report reference
polarizabilities and $C_6$ coefficients for free-standing
atoms and ions (for ions with $N\leq Z+1$). However,
polarizabilities are often desired for their
utility in embedded atom/ion models - such as for high-level
calculations using van der Waals dispersion corrections
(e.g. \rcites{Grimme2004,Grimme2006,Grimme2010,
Tkatchenko2009,Tkatchenko2012,DiStasio2012}),
or for (semi-)classical force-field models. We thus
also report a slightly modified set of reference coefficients
for this purpose.

Atoms and ions ``embedded'' in a larger system such as a molecule
or material can behave very differently to their free-standing
counterparts.
Most notably, the surrounding environment of an anion has a
substantial effect on the behaviour of its outermost electron(s).
They are only very weakly bound in the
free-standing case, with an asymptotic effective potential
going to zero as $~r^{-3}$. When embedded, the other electrons and
nucleii will introduce an effective confining potential, which
can be approximated [see e.g. \rcite{Cruz2009-1-Patel}]
by a positve power law function of $r$ such as
$(r/r_a)^{\sigma_a}$ where $r_a$ is an effective embedding radius
and $\sigma_a$ governs the sharpness. Clearly the energetics
of the outermost orbital and unoccupied orbitals
are very different in both cases.

But the same electrons that are most sensitive to the embedding
environment, namely electrons in the outermost electronic shell(s),
are the ones that contribute the most to the polarizibilty.
Thus the polarizability of an embedded system is highly sensitive
to its enviroment and care must be taken in considering what
``anions'' should be used in embedding theories.
This is especially pertinent for open shell systems which are
likely to be the most sensitive to the environment.

To account for embedding, we thus approximate embedded
anions using a minimal chemistry model chosen to ensure
that the ``free'' ions behave as
closely as possible to their embedded counterparts without taking
into account the full details of the chemical environment.
To this end, we carry out polarizability calculations using a
frozen orbital model, in which the polarizabilities
$\alpha\equiv \alpha[v_s[v_Z],\{f_i\}]$ are treated (in TDDFT)
as a functional of the \emph{interacting potential} $v_s$
via $\chi[v_s,\{f_i\},f_{\Hxc}](\vr,\vr';i\omega)$,
discussed in detail just before Sec.~\ref{sec:converge}.
This allows us to introduce some embedding effects
via our choice of $v_s$ arising from external potential $v_Z=-Z/r$.

Our minimal chemistry model involves determining $v_s$
from the following rules:
\begin{enumerate}
\item For neutral atoms and cations, find the atomic potential
  self-consistently using LEXX theory, Hund's rules and the
  aufbau principle,
  (see Appendix~\ref{app:TM} for further details on Rows 4-6)
  then rescale the polarizabilities using \eqref{eqn:rescaling};
\item For single anions A$^-$ of halides
  ($N=10,18,36,54,86$, $Z=N-1$) calculate
  the ionic potential self-consistently using LEXX theory
  and then rescale (like for neutral atoms);
\item For the remaining single anions A$^-$ use
  the self-consistent LEXX potential of the neutral atom
  and do not rescale;
\item For all double anions A$^{2-}$ use
  the self-consistent LEXX potential of the neutral atom
  and do not rescale;
\end{enumerate}
This method produces an alternative set of frequency dependent
anionic polarizabilities that, we feel, may better reflect
the reality of embedded anions.
This alternative data set is provided in the
supplementary data\cite{Supp}.

\section{Two-Lorentzian polarizability model}
\label{app:TwoLor}

From equation~\eqref{eqn:alphaFormal}
\begin{align}
  \alpha(i\omega)\approx& \sum_{kl}
  \frac{d_{kl}}{(E_k-E_l)^2+\omega^2}
\end{align}
it is clear that the imaginary
frequency dependence of $\alpha$ is a sum over Lorentzian
functions $d/(\omega^2+\Omega^2)$. In a typical atom
one Lorentzian will typically have a denominator
(the square of the lowest excitation energy + $\omega^2$) that is
significantly smaller than the other terms for small $\omega$,
and thus dominates the $C_6$ coefficient.
When combined with the known limit $\lim_{\omega\to\infty}
\alpha(i\omega)=N/\omega^2$ (i.e. $\sum d_{kl}=N$)
this suggests that a reduced
number of Lorentzians should be sufficient to represent
the imaginary frequency dipole polarizabilities.

In fact, our results suggest that $\alpha(i\omega)$ for atoms/ions
can be approximated by just two Lorentzians with minimal
loss of accuracy.
Thus, for every atom and ion with nuclear charge $Z$ and
electron number $N$ we write
\begin{align}
  \alpha_X(i\omega)\approx&
  \sum_{c=1,2} \frac{a_{c,X}}{\omega^2 + \Omega_{c,X}^2}
  \label{eqn:alphaModelApp}
\end{align}
without any great loss of accuracy. Here the parameters $a_1$,
$\Omega_1$ and $\Omega_2$ are varied to minimze
$\int (\alpha-\alpha^{\text{Numeric}})^2 d\omega$
while $a_2=N-a_1$ is kept fixed to ensure
that the polarizability has the correct asymptote.

We note that Figari~\latin{et~al}\cite{Figari2007-TCA,Figari2007-CL}
have carefully studied similar pseudospectral methods to the one
employed here. They showed that four Lorentzians are generally
sufficient for very high accuracy and that
careful treatment of the frequencies $\Omega_c$
can further improve results. However, given the merely
``moderate'' quality of our inputs, we feel that such
an analysis would not offer meaningful benefits for our
present work.

%\bibliography{ACFDT,vanDerWaals,DFT,Wannier,Misc,Experiment,Hybrid,%
%QMGeneral,ISTLS,Frac,OEP,confinement,Graphene,mbdvdw}
\providecommand{\latin}[1]{#1}
\providecommand*\mcitethebibliography{\thebibliography}
\csname @ifundefined\endcsname{endmcitethebibliography}
  {\let\endmcitethebibliography\endthebibliography}{}

\end{document}